\documentclass[aps,prd,groupedaddress]{revtex4}

\newcommand{\tr}{{\mathrm{tr}}}
\newcommand{\TR}{{\mathrm{Tr}}}

\begin{document}

LPT-Orsay-02/73; IHES/P/02/51

\title{Variational analysis of the deconfinement phase transition}

\author{Ian Kogan$^{1,2,3}$, Alex Kovner$^4$, and J. Guilherme Milhano$^5$\\ }
\affiliation{$^1$ Theoretical Physics, Oxford University, 1 Keble
road, Oxford OX1 3NP, UK;\\
$^2$  IHES, 35 route de Chartres,
91440, Bures-sur-Yvette,  France \\
$^3$ Laboratoire de Physique Th\'eorique,
Universit\'e de Paris XI, 91405 Orsay C\'edex, France\\
 $^4$ Department of Mathematics and
Statistics, University of Plymouth, 3 Kirkby place, Plymouth PL4
8AA, UK,\\
 $^5$ Department of Theoretical Physics, Faculty of Sciences, Vrije Universteit,
De Boelelaan 1081, NL-1081 HV Amsterdam, The Netherlands }

\begin{abstract}
We study the deconfining phase transition in 3+1 dimensional pure
$SU(N)$ Yang-Mills theory using a gauge invariant variational
calculation. We generalize the variational ansatz of \cite{kk} to
mixed states (density matrices) and minimize the free energy. For
$N\ge 3$ we find a first order phase transition with the
transition temperature of $\mathsf{T_{c}}\simeq 450{\rm Mev}$.
Below $\mathsf{T_{c}}$ the Polyakov loop has vanishing expectation
value, while above $\mathsf{T_{c}}$, its average value is nonzero.
According to the standard lore this corresponds to the deconfining
transition. Within the accuracy of our approximation the entropy
of the system in the low temperature phase vanishes. The latent
heat is not small but, rather, is of the order of the
nonperturbative vacuum energy.
\end{abstract}

\maketitle

\section{Introduction}
Attempts to understand the nature of the deconfining phase
transition in QCD date back almost 25 years. Since the pioneering
work of Polyakov \cite{polyakov}, much effort has been made to
study the basic physics as well as the quantitative
characteristics of the transition. The high temperature phase of
QCD is widely believed to resemble an almost free plasma of quarks
and gluons. At asymptotically high temperatures this is confirmed
by explicit perturbative calculations of the free energy
\cite{perturb}. Perturbation theory in its simplest form, however,
is valid only at unrealistically high temperatures. In recent
years a different promising avenue has been explored. This
incorporates analytical resummation of the effects of the gluon
screening mass into the 3D effective Lagrangian, which is then
solved numerically by 3D lattice gauge theory methods
\cite{nieto,mikko}. The results of this approach seem to be in
good agreement \cite{hart} with direct 4D lattice gauge theory
calculations \cite{karsch} all the way down to $2\mathsf{T_{c}}$.
The numerical results indicate that, interestingly enough,
although asymptotically the free energy does approach that of the
free partonic plasma, the deviations from the Stefan-Boltzmann law
even at temperatures of order $10 \mathsf{T_{c}}$ are quite
sizable, of order of $15\%$. One can interpret this as an
indication that the interesting physics of the transition region
remains important even at these high temperatures.

Although we are quite advanced in the understanding of the high
temperature phase, the transition region itself is understood very
poorly. This region of temperatures
$\mathsf{T_{c}}<\mathsf{T}<2\mathsf{T_{c}}$, is of course the most
interesting one, since it is in this region that the transition
between ``hadronic" and ``partonic" degrees of freedom occurs. In
fact, it is quite possible that the dynamics of the transition
region is dominated by the $Z_N$ magnetic vortices \cite{thooft},
which are responsible for the abrupt increase in entropy
\cite{kaks}. The study of the transition region is a complicated
and inherently nonperturbative problem which has not been tackled
so far by analytical methods.

The purpose of this paper is to explore an analytical
investigation of the phase transition region. Since the nature of
the problem is nonperturbative, a nonperturbative method is called
for. Unfortunately there are not many of those on the market. It
does however seem to us that the variational method developed
several years ago in \cite{kk} is quite suitable. We are
encouraged by the fact that the variational approximation of
\cite{kk} manages to reproduce all known infrared features in
lower dimensional confining theories \cite{kk1,kk2,ks,brogakoko1},
and also gives a stable picture of instanton dynamics in QCD
\cite{brogakoko}. Even though, as any variational method, it has
an element of guesswork about it, it is a first principle
calculation. For such a difficult physics problem as the one at
hand, clearly even an exploratory investigation which provides a
qualitative picture should be very valuable and is well worth
attempting.

In this paper, therefore, we will study the deconfining phase
transition in a pure $SU(N)$ Yang-Mills theory using the
variational approach of \cite{kk} suitably extended to finite
temperature. We will minimize the relevant thermodynamic potential
at finite temperature, i.e. the Helmholtz free energy, on a set of
gauge invariant density matrices.

The results of our calculation are very interesting. We find that
the phase transition in gluodynamics is closely related to the
symmetry breaking phase transition in the effective $\sigma$-model
for the Polyakov loop variable $U$. This $\sigma$-model arises
explicitly in our calculation, and its parameters depend on the
variational parameters of the trial density matrix. At low
temperature the effective $\sigma$-model is in the disordered
phase, and $\langle U\rangle=0$. The entropy of the Yang-Mills
thermal density matrix vanishes in the low temperature phase.  At
the critical temperature of order $\mathsf{T_{c}}\simeq 450 {\rm
Mev}$, the best variational density matrix changes abruptly, and
above $\mathsf{T_{c}}$ corresponds to the ordered phase of the
$\sigma$-model. Thus at $\mathsf{T}>\mathsf{T_{c}}$ we find
$\langle U\rangle\ne 0$. The transition itself turns out to be
strongly first order for large $N$. This is consistent with recent
lattice gauge theory calculations \cite{mike}.  These results are
not surprising in themselves.  They are expected from the general
relation between the average value of the Polyakov loop and the
(de)confining properties of the density matrix. Moreover, the
order of the transition is predicted by the Svetitsky-Yaffe
conjecture \cite{ben}. The vanishing of the entropy in the
confining phase is also very natural, since the glueballs are
heavy and their contribution to the entropy should be suppressed by
the Boltzmann factor $\exp[-M_g/T]$. Nevertheless, we are not
aware of any first principle analytic calculation in QCD which
provides an explicit realization of these general arguments.

This paper is structured as follows. In sec.~2 we recap the
variational ansatz of \cite{kk} for energy minimization at zero
temperature and recall the  methods of evaluating the expectation
values in the variational wave function as well as the results of
the minimization procedure.
 In sec.~3 we
generalize the variational ansatz so that it incorporates not only
pure states but also general density matrices. We introduce the
two variational parameters with respect to which the free energy
is to be minimized. In sec.~4 we recast the problem of the
minimization of free energy in terms of an effective nonlinear
$\sigma$-model. In sec.~5 we calculate the Helmholtz free energy
of the trial density matrix and perform the minimization. We
conclude in sec.~6 by discussing our results.

\section{The Variational Calculation at Zero Temperature}
The variational method proposed in \cite{kk} is based on
minimization of the expectation value of the Hamiltonian on a set
of gauge invariant states. The set of states is chosen so that
their functional form is a relatively simple generalization of the
vacuum of the free theory. Ideally, one would like to choose a set
of variational states that is both simple enough to perform
explicit calculations and rich enough to allow for variations of
its parameters to span interesting physics. One starts then with a
Gaussian state
\begin{eqnarray}
\Psi_0[A_i^a]=\exp\left\{ - \frac{1}{2}\int d^{3}x d^{3}y
 A_i^a(x)
 (G^{-1})^{ab}_{ij}(x,y)
 A_j^b(y)\right\}\, ,
\end{eqnarray}
where the set of functions  $G^{ij}_{ab}(x)$ are variational
parameters. Expectation values of various operators are easily
calculable in this set of states. One also expects that the
freedom allowed by variation of $G$ is wide enough to probe the
nonperturbative physics of confinement. The problem is that these
states are not gauge invariant and, therefore, as such do not
belong to the physical Hilbert space of gluodynamics. To remedy
this problem one projects the states onto the gauge invariant
subspace by gauge transforming them and integrating over the whole
gauge group:
\begin{equation}\Psi[A_i^a]=\int DU(x)
 \exp\left\{-{1\over 2}\int d^{3}x d^{3}y
\ A_i^{Ua}(x)G^{-1ab}_{ij}(x-y)\ A_j^{Ub}(y)\right\}
\label{an}
\end{equation}
with $A_i^{Ua}$ defined as
\begin{equation} \ A^{Ua}_i(x)=S^{ab}(x)A_i^b(x)+
\lambda_i^a(x)\, ,
\label{gt}
\end{equation}
and
\begin{eqnarray}
S^{ab}(x)={1\over 2}tr\left(\tau^aU^\dagger\tau^bU\right);
\ \ \ \lambda_i^a(x)={i\over g}tr\left(\tau^aU^\dagger
\partial_iU\right)
\label{defin}\end{eqnarray} with $\tau$ --- traceless $
N\times N$ hermitian matrices --- normalized by
\begin{equation}
    \tr (\tau^a \tau^b) = 2 \delta^{ab}\, ,
\end{equation}
and obeying the completeness condition for $SU(N)$
\begin{equation}
    \label{eq:fierz}
    \tau_{ij}^{a} \tau_{kl}^{a} = 2 \big( \delta_{il}\delta_{jk}-\frac{1}{N}
    \delta_{ij}\delta_{kl}\big)\, .
\end{equation}

The integration in eq.~(\ref{an}) is performed over the space of
special unitary matrices with the $SU(N)$ group invariant measure.
This integration projects the original Gaussian state onto a
colour singlet. Due to the projection operation, the calculation
of expectation values in this state is much more involved than in the case of 
a simple gaussian. Thus it is a very difficult task to perform the
full functional minimization with respect to $G^{ij}_{ab}(x)$.
This can be done and was indeed performed in simpler calculations
in 2+1 dimensional models \cite{kk1,kk2}. In these models the
approach was also extended to provide for the calculation of the
confining potential between the external charges \cite{ks}. It has
also been applied to 1+1 dimensional models with fermions
\cite{brogakoko1}. In these simple models the variational
approximation has performed very well, reproducing all known
results.

In nonabelian theories, unfortunately, full functional
minimization is beyond our calculational abilities. Therefore to
make headway, the width of the Gaussian $G$ was restricted in
\cite{kk} to a one parameter form as follows
\begin{equation}
G^{ab}_{ij}(x-y)=\delta^{ab}\delta_{ij}G(x-y)
\label{an1}
\end{equation}
with the Fourier transform of $G$ taken to be
\begin{eqnarray}
\label{G}
 G^{-1}(k) = \left\{ \begin{array}{ll} \sqrt{ k^{2} ~} &
\mbox{ if  $ k^2>M^2$}\\ M &  \mbox{ if $k^2<M^2$}
\end{array}
\right.\, .
\end{eqnarray}
The dimensional parameter $M$ is thus the
only variational parameter in the calculation.

The expectation value of any gauge invariant operator $O$ in the
variational state eq.~(\ref{an}) is given by the following integral
\begin{eqnarray}
\langle O\rangle&=& {1\over Z}\int DU \langle O\rangle_{A}\, ,\\
\langle O\rangle_{A}&=& \int DA e^{-{1\over 2}\int dx dy A_i^{Ua}(x)
G^{-1}(x-y) A_i^{Ua}(y) }O e^{-{1\over 2}\int dx' dy'
A_j^b(x')G^{-1}(x'-y') A_j^b(y')}\nonumber\, .
\label{expect}
\end{eqnarray}
Note that only one group integral is present. This is because for
a gauge invariant operator $O$ only one of the states has to be
gauge projected.

Since the gauge transform of a vector potential is a linear
function of $A$ eq.~(\ref{gt}), for fixed $U(x)$ this is a Gaussian
integral, and can therefore be performed explicitly for any
reasonable operator $O$. The nontrivial part of the calculation is
the  path integral over the group variable $U(x)$. Consider first
the normalization factor $Z$. After integrating over the vector
potential we obtain:
\begin{equation}
Z=\int DU\exp\{-{\cal S}[U]\} \label{sigma}
\end{equation}
with the action
\begin{equation}
{\cal S}[U]={1\over 2} Tr \ln{\cal M} +{1\over
2}\lambda[G+SGS^T]^{-1}\lambda \, .\label{action}
\end{equation}
Here
\begin{equation}
S^{ab}_{ij}(x,y)=S^{ab}(x)\delta_{ij}\delta(x-y), \ \ {\cal
M}^{ab}_{ij}(x,y)=
[S^{Tac}(x)S^{cb}(y)+\delta^{ab}]G^{-1}(x-y)\delta_{ij}\, ,
\label{def1}
\end{equation}
and we use matrix product notation, which implies summation over
the colour and Lorentz indices, as well as integration over the
spatial coordinates. In the rest of this paper ${\rm Tr}$ stands
for summation over all indices, including the space coordinates,
while ${\rm tr}$ stands for summation over the colour indices
alone. The path integral eq.~(\ref{sigma}) defines a partition
function of a nonlinear $\sigma$-model with the target space
$SU(N)/Z_N$ in three dimensional Euclidean space. The fact that
the target space is $SU(N)/Z_N$ rather than $SU(N)$, follows from
the observation that the action eq.~(\ref{action}) is invariant
under local transformations belonging to the centre of $SU(N)$.
This can be trivially traced back to invariance of $A_i^a$ under
gauge transformations that belong to the centre of the gauge
group.

The matrix $U(x)$ has a well defined gauge invariant meaning. It
arises in eq.~(\ref{expect}) as a relative gauge transformation
between the bra and the ket gaussian wave functionals. As such it
plays exactly the same role as Polyakov's loop $P$ at finite
temperature.
 We will therefore refer to it as the Polyakov loop in
the following \cite{note1}.
 The contribution of a given $U(x)$ to the partition function
eq.~(\ref{sigma}) and to other expectation values corresponds to
the contribution due to the off diagonal matrix element between
the initial Gaussian and the Gaussian gauge rotated by $U(x)$.
Therefore, if matrices $U(x)$ which are far from unity give a
significant contribution to the partition function, it means that
the off diagonal contributions are large and gauge projection is
indeed mandatory to get physically sensible results.

The action of this $\sigma$-model is rather complicated.
 It is a nonlocal and a nonpolynomial functional of $U(x)$.
 The following set of approximations was employed in \cite{kk} to
 estimate this path integral.
 First, since  the bare coupling constant of the Yang
Mills theory is small and it also enters as a coupling constant of
the nonlinear $\sigma$-model, the high momentum modes of the field
$U$ are integrated out perturbatively. The result of the
perturbative one loop integration of the modes with spatial
momenta $k>M$ is a much simpler $\sigma$-model which involves low
momentum modes only. The action of this $\sigma$-model is
\begin{equation}
{\cal S}_L[U] ={M \over 2g^2(M)} tr \int d^{3}x ~
\partial_iU^\dagger(x)\partial_iU(x)\, ,
\label{lowm}
\end{equation}
where $M$ also serves as the ultraviolet cutoff. To be more
precise, due to the $Z_N$ local symmetry of the original theory
eq.~(\ref{action}), the action for the low momentum modes is
slightly different. The derivatives should be understood as $Z_N$
covariant derivatives. The most convenient way to write this
action, would be to understand $U(x)$ as belonging to $U(N)$
rather than $SU(N)$, and to introduce a $U(1)$ gauge field by
\begin{equation}
{\cal S}_L={1\over 2}{M \over g^2} tr
 \int d^{3}x ~(\partial_i-iA_i)U^\dagger(x)(\partial_i+iA_i)U(x)\, .
\end{equation}
This defines a $\sigma$-model on the target space $U(N)/U(1)$,
which is isomorphic to $SU(N)/Z_N$. However, since the difference
between the two actions is not important at large number of
colours, we will disregard this subtlety in the following and use
eq.~(\ref{lowm}).

As a result of the running of the QCD coupling constant, the
effective coupling $g^2(M)$ is given by the one loop QCD formula
\cite{brko}. Clearly for the perturbative integration to be
consistent we should find that the energy is eventually minimized
for high enough value of the scale $M$, so that $\alpha_s(M)\ll
1$. The actual value found in \cite{kk} is $\alpha_s(M)\simeq
.25$, which is indeed reasonably small. To solve the resulting low
energy $\sigma$-model the mean field approximation and
perturbation theory are used. We will describe this procedure here
in some detail, since we will use it in later sections to analyze
the finite temperature case.

The Hamiltonian of the pure Yang-Mills theory is
\begin{equation}
{\mathsf H}= \int d^{3}x \left[{1\over 2}E^{a2}_i+{1\over
2}B^{a2}_i\right]\, , \label{ham}
\end{equation}
where
\begin{eqnarray}
E^a_i(x)&=&i{\delta\over \delta A^a_i(x)}\, , \nonumber \\
B^a_i(x)&=&{1\over 2}\epsilon_{ijk}
\{\partial_jA_k^a(x)-\partial_kA^a_j(x)+gf^{abc}A_j^b(x)A_k^c(x)\}\, .
\end{eqnarray}

To calculate the expectation value of ${\mathsf H}$ in the variational state
we first perform the integral over the gauge fields $A$ at fixed
value of the Polyakov loop $U$. The result of this integration
(for details see \cite{kk}) for the energy density is
\begin{eqnarray}
{\langle 2{\mathsf H}\rangle\over V} &=&{3(N^2-1)\over 2} G^{-1}(x,x) +
(N^2-1) \partial_i^x\partial_i^yG(x-y)|_{x=y} \nonumber \\
&-&{1\over 4 V}\int d^{3}xd^{3}y\langle\lambda_{iH}^a(x)G^{-2}(x-y)
\lambda_{iH}^a(y)\rangle_U +{1\over 4} \langle(\epsilon_{ijk}\partial_j
\lambda^a_{kH})^2\rangle_U
\nonumber \\
&-&{M^{2}\over 4 V}\int d^{3}x
\langle\lambda_{iL}^a(x)\lambda_{iL}^a(x)\rangle_U \, ,\label{energy}
\end{eqnarray}
 where  the averaging  over the $U$ field
should be performed with the $\sigma$-model action, eq.~(\ref{action}). As
the high momentum modes are only considered to one loop order, the
$\sigma$-model action reduces to
\begin{equation}
{\cal S} ={1\over 4}\int dx dy \lambda^a_{iH}(x)G^{-1}(x-y)
\lambda^a_{iH}(y)+ {M \over 2g^2(M)} tr \int d^{3}x ~
\partial_iU_L^\dagger(x)\partial_iU_L(x) \, .\label{hl}
\end{equation}
The subscripts $H$ and $L$ indicate that the corresponding
expressions contain only high and low momentum components of the
fields respectively. Integration over the high momentum modes
gives
\begin{equation}
{2\langle{\mathsf H}\rangle \over V} = {N^2-1\over 10\pi^2}M^4-{M^{2}\over 4
V}\int d^{3}x \langle\lambda_{iL}^a(x)\lambda_{iL}^a(x)\rangle_U\, ,
\end{equation}
where now only the contribution of the low momentum modes remains
to be evaluated.

This last step of the calculation is the most difficult one. The
nonlinear $\sigma$-model eq.~(\ref{lowm}) possesses a global
$SU_L(N)\otimes SU_R(N)$ symmetry. Thus it has a symmetry
restoring phase transition as a function of $M$. The coupling
constant $g^2(M)$ enters this action in the same way as
temperature enters in the classical statistical problem of the
$SU(N)$ spin system. High values of $M$ correspond to low values
of $g^2(M)$ and thus to the low temperature regime in the
$\sigma$-model. In this regime the global $SU_L(N)\otimes SU_R(N)$
symmetry is spontaneously broken down to the vector subgroup
$SU_V(N)$. The model itself is weakly interacting and thus one can
use perturbation theory. However, it is well known that the model
undergoes a phase transition from the ordered low temperature
phase to the disordered phase at some critical temperature
$g^2_c$. The disordered phase is clearly nonperturbative. To treat
it quantitatively the mean field approximation was used in
\cite{kk}. The partition function of the $\sigma$-model is
rewritten by introducing a (hermitian matrix) auxiliary field
$\alpha$ which imposes a unitarity constraint on $U(x)$
\begin{eqnarray}
Z&=&\int DU D\alpha\exp\left(- {\cal S}[U,\alpha]\right)\, ,
\\
{\cal S}[U,\alpha] &=& {M\over 2g^2(M)}
 tr \int d^{3}x \left[\partial_i
U^\dagger(x)\partial_iU(x) + \alpha \left(U^\dagger
U-1\right)\right] \, .\nonumber
\end{eqnarray}
In this step we have disregarded the fact that $U$ is an $SU(N)$
matrix rather than $U(N)$. Thus, strictly speaking, our
calculation applies only at large $N$.
 The mean field equations for this action are
\begin{equation}
\langle U^\dagger U\rangle=1 \, ,\label{one}
\end{equation}
\begin{equation}
\langle\alpha U\rangle=0 \, .\label{two}
\end{equation}
From eq.~(\ref{two}) it follows that either $\langle\alpha\rangle=0$,
$\langle U\rangle\ne 0$ (the ordered, broken symmetry phase with massless
Goldstone bosons),  or $\langle\alpha\rangle\ne0$, $\langle U\rangle= 0$ (the disordered,
unbroken phase with massive excitations).

In the disordered phase the expectation value of $\alpha$ is
proportional to the unit matrix
\begin{equation}
\langle\alpha_{\alpha\beta}\rangle=\alpha^2 1_{\alpha\beta}\, .
\end{equation}
Eq.~(\ref{one}) then becomes
\begin{equation}
2N^2{g^2(M)\over M}\int_0^M{d^3k\over (2\pi)^3}{1\over
k^2+\alpha^2}= {N^{2} g^2(M)\over \pi^2} \left(1 - {\alpha\over M}
\arctan{M\over \alpha}\right) = N \, .\label{gap}
\end{equation}
The gap equation, eq.~(\ref{gap}), has a solution only for couplings
 $g^{2}(M)$
 larger than the
 critical coupling $g^{2}_{c}$, which
is determined by the condition that $\alpha = 0$
\begin{equation}
\alpha_s^c={g^2_c\over 4\pi}={\pi \over 4}{1\over N}\, . \label{gc}
\end{equation}
The low momentum mode contribution to the ground state energy is
\begin{equation}
N^2M\int_0^M{d^3k\over (2\pi)^3}{k^2\over k^2+\alpha^2}= {N^2\over
2\pi^2}M\left[{1\over 3}M^3-\alpha^2M+\alpha^3 \arctan{M\over
\alpha}\right]\, .
\end{equation}

The final mean field expression for the ground state energy
density is
\begin{equation}
{\mathsf E}= {N^2\over 4\pi^2}M^4\left[-{2\over 15}+{\alpha^2\over M^2}
{\alpha_s^c\over \alpha_s(M)}\right]\, , \label{fen}
\end{equation}
where $\alpha_s(M)$ is the QCD coupling at the scale $M$,
$\alpha_s^c$ is given by eq.~(\ref{gc}), and $\alpha$ is
determined by
\begin{equation}
{\alpha\over M}\arctan {M\over
\alpha}={\alpha_s(M)-\alpha_s^c\over \alpha_s(M)}\, .
\end{equation}
This function has the minimum at the critical point
$\alpha_s(M)=\alpha_s^c$. With the one loop Yang-Mills $\beta$
function and $\Lambda_{QCD}=150 {\rm Mev}$, we find for $N=3$
\begin{equation}
M_c=\Lambda_{QCD}e^{24\over 11}=8.86\Lambda_{QCD}=1.33 \mathrm{Gev}\, .
\label{mc}
\end{equation}

Eq.~(\ref{fen}) is valid only in the disordered phase. Here we
expect the mean field approximation to be reasonably good, since
the spectrum of the $\sigma$-model is massive and fluctuations are
not expected to be overwhelmingly large.

In the ordered phase, on the other hand, the situation is
different. The spectrum contains $O(N^2)$ ``Goldstone bosons". In
the mean field approximation in addition it also has $O(N^2)$
massive particles --- the analogs of $\delta$ particles in QCD. In
the mean field the $\delta$'s are all stable. In reality of
course, since there is nothing to prevent them from decaying into
the Goldstone bosons, they are not at all stable. In fact, the
further into the ordered phase one gets, the less stable they are
due to the increase of the phase space available for decay. Thus
we expect the mean field to be unreliable in the ordered phase.
Asymptotically at very large $M$ the mass of $\delta$ becomes much
higher than the ultraviolet cutoff of the $\sigma$-model. For
these values of $M$ the massive modes decouple even within the
mean field approximation, and one recovers the perturbative
result. This decoupling, however, occurs extremely slowly. We have
checked that the mean field formulae tend to the perturbative ones
only when the mass of the would be stable massive states is about
1000 times the ultraviolet cutoff. This is, of course, unphysical,
confirming our suspicion that the mean field approximation cannot
be used on the ordered side of the transition.

The simplest option to calculate the contribution of the low
momentum modes to the energy in the ordered phase is to use
perturbation theory. The {\it raison d'etre} for this is the
following. Perturbation theory is certainly appropriate for large
enough values of $M$, where the expectation value of the $U$ field
is of order unity. On the other hand, we know from the numerical
studies of \cite{kogut} that the phase transition in our
$\sigma$-model is strongly first order. More specifically,
according to \cite{kogut} the transition occurs when the
expectation value of $U$ is greater than $.5$. We thus expect
perturbation theory to be qualitatively reliable all the way down
to the transition point. Since we have neglected previously many
perturbative contributions of order $g^2$, in order to be
consistent we have to limit ourselves to the leading order
contribution only. Calculating with this precision the energy of
the trial states for $M\ge M_c$ we get
\begin{equation}
{{\mathsf E}(M)\over V}= {N^2-1\over 120\pi^2}M^4\, . \label{largeM}
\end{equation}
Thus for $M\ge M_c$ the energy is a monotonically increasing
function of $M$.

Eq.~(\ref{mc}) is thus the result of the minimization of the energy
over the whole range of the variational parameter $M$.

\section{The variational Ansatz for the density matrix}
In order to extend the variational analysis to finite temperature
we have to generalize our ansatz so that it includes mixed states.
In scalar theories the Gaussian approximation has a long history
of applications at finite temperature \cite{eboli,moshe}.
We generalize our ansatz along the same lines.

We start by considering the density matrices which in the field
basis have Gaussian matrix elements
\begin{equation}
\label{eq:iniansatz}
    \varrho[A,A']  = \exp \Bigg\{ -\frac{1}{2} \int_{x,y}
            A^{a}_{i}(x)G^{-1ab}_{ij} (x,y)A^{b}_{j}(y)
            +A^{'a}_{i}(x)G^{-1ab}_{ij} (x,y)A^{'b}_{j}(y)
-2A^{a}_{i}(x)H^{ab}_{ij} (x,y)A^{'b}_{j}(y)\Bigg\}\, .
\end{equation}

As before, we take the variational functions  diagonal in both
colour and Lorentz indices, and translationally invariant
\begin{eqnarray}
    G^{-1ab}_{ij} (x,y) &= &\delta^{ab} \delta_{ij}
    G^{-1}(x-y)\, ,\\
H^{ab}_{ij} (x,y) &= &\delta^{ab} \delta_{ij}
    H(x-y)\, .\nonumber
\end{eqnarray}
Then
\begin{equation}
    \tilde\varrho[A,A']
    = \exp \bigg\{ -\frac{1}{2} \int_{x, y}
    \bigl(A G^{-1} A +  {A'} G^{-1} {A'}
    -2A H {A'} \bigl) \bigg\}\, .
    \label{tilderho}
\end{equation}
Note that for $H=0$ this density matrix represents a pure state,
since it can be written in the form
\begin{equation}
    \tilde\varrho=|\Psi[A]><\Psi[A]|
\end{equation}
with $\Psi[A]$ a gaussian wave function. At nonzero $H$ the
density matrix is, however, mixed. The magnitude of $H$,
therefore, determines the entropy of this trial density matrix.

We now make an additional simplification in our ansatz. First, we
restrict the functions $G^{-1}(x)$ to the same functional form as
at zero temperature, eq.~(\ref{G}). Further, we will take $H(k)$ to
be small and nonvanishing only at low momenta
\begin{eqnarray}
\label{H}
 H(k) = \left\{ \begin{array}{ll} 0  &
\mbox{ if  $ k^2>M^2$}\\ H \ll M&  \mbox{ if $k^2<M^2$}
\end{array}
\right.\, .
\end{eqnarray}

The logic behind this choice of ansatz is the following. At finite
temperature we expect $H(k)$ to be roughly proportional to the
Bolzmann factor $\exp\{-{\mathsf E}(k)\beta\}$. In our ansatz, the role of
one particle energy is played by the variational function
$G^{-1}(k)$. We will be  interested only in temperatures close to
the phase transition, and those we anticipate to be small,
$\mathsf{T_{c}}\le M$. For those temperatures one particle modes
with momenta $k\ge M$ are not populated, and we thus put $H(k)=0$.
For $k\le M$ the Bolzmann factor is nonvanishing, but small.
Further, it depends only very weakly on the value of the momentum.
We will have, of course, to verify a posteriori that our
assumptions about the smallness of $\mathsf{T_{c}}$ and $H$ are
justified. As we will see later, this turns out indeed to be the
case with reasonable accuracy.

As before, we  explicitly impose gauge invariance by projecting
$\varrho$ onto a gauge invariant sector
\begin{equation}
    \label{eq:ginvansatz}
    \varrho[A,A'] = \int {\cal D}U'{\cal D}U''  \exp \bigg\{
    -\frac{1}{2} \int_{x,y}
    A^{U'} G^{-1} A^{U'} +  {A'}^{U''} G^{-1}{A'}^{U''}
    - 2A^{U'} H {A'}^{U''}
    \bigg\}\, .
\end{equation}

One of the group integrations in eq.~(\ref{eq:ginvansatz}) is
redundant, since we will only calculate the quantities of the form
$\TR\varrho O$, with $O$ being gauge invariant. We thus have the
following ansatz for the density matrix
\begin{equation}
    \label{eq:ansatz}
    \varrho[A,A'] = \int {\cal D}U \exp \bigg\{
    -\frac{1}{2} \int_x
    A G^{-1} A +  {A'}^U G^{-1}{A'}^U - 2A H {A'}^U
    \bigg\}\, .
\end{equation}

This expression is not explicitly normalized to unity.
Nevertheless, we find it convenient to refer to it as density
matrix while explicitly inserting a normalization factor whenever
necessary. Thus the average of a gauge invariant operator ${\cal
O}$ is given by
\begin{eqnarray}
\label{eq:expvalue}
    \langle {\cal O}\rangle_{A,U}
    & =& Z^{-1}\TR (\varrho {\cal O}) \nonumber  \\
    &=& Z^{-1} \int   {\cal D}U{\cal D}A
    \:{\cal O}(A,A')
    \exp \bigg\{ -\frac{1}{2} \int_x
    A G^{-1} A +  {A'}^U G^{-1}{A'}^U
    - 2A H {A'}^U \bigg\}\Bigg|_{A'=A}\, ,
\end{eqnarray}
where $Z$ is the normalization of the trial density matrix
$\varrho$, i.e.
\begin{eqnarray}
    Z&=&\TR\varrho = \int {\cal D}U{\cal D}A \exp \bigg\{
    -\frac{1}{2} \int_x
    A G^{-1} A +  {A}^U G^{-1}{A}^U
    -2 A H {A}^U \bigg\} \nonumber\\
    &=&\int {\cal D}U{\cal D}{\tilde A} \exp \biggl\{
    -\frac{1}{2} \int_x
    {\tilde A}\Delta{\tilde A}
    +\lambda \bigl(
    G^{-1} - \omega \Delta^{-1} \omega^{T} \bigr)
    \lambda\biggr\}
\end{eqnarray}
with
\begin{eqnarray}
    {\tilde A} &=& A + \lambda \omega\Delta^{-1}\, , \\
    \Delta &=& 2 G^{-1} \Bigl(1-\frac{HG}{2} (S+S^T)\Bigr)\, , \\
    \omega &=& (G^{-1}S-H)\, .
\end{eqnarray}

The ${\tilde A}$ integration can then be easily performed to yield
\begin{equation}
    \TR\varrho = \int {\cal D}U
    \exp \bigg\{
    -\frac{1}{2}
    \lambda \bigl(
    G^{-1} - \omega \Delta^{-1} \omega^{T}
    \bigr) \lambda
    -\frac{3}{2}\TR\ln\frac{\Delta}{2}\bigg\}\, .
\end{equation}

We now adopt the same strategy for treating the high momentum
modes of $U$ as at $\mathsf{T}=0$. Namely, they are integrated
perturbatively to one loop accuracy. The result is the effective
$\sigma$-model for the matrices $U$ with momenta below $M$. The
coupling constant $g$ of this $\sigma$-model gets renormalized
as before according to the one loop Yang-Mills $\beta$-function,
and thus has to be understood as $g(M)$. Additionally, due to
independence of $H$ on momentum, for low momentum modes of $U$ the
function $H(x-y)$ is equivalent to $H\delta^3(x-y)$.

The final approximation has to do with the fact that $H$ is
assumed to be small. For arbitrarily large $H$ the variational
calculation is forbiddingly complicated even with all the above
mentioned simplifications. This is because the gauge projection
renders the calculation of entropy in the general case unfeasible.
However, at small $H$ we only need to calculate the leading term
in entropy. This calculation can indeed be done, and is described
in the next section.  Since we are only calculating the leading
order contribution in $H$, we only have to consider corrections to
the $\sigma$-model action of first order in $H$. With this in
mind, the normalization factor becomes
\begin{equation}
    \label{eq:trvarrho}
    \TR\varrho = \int {\cal D}U \exp \bigg\{
    -\frac{1}{2}\lambda\Big(
    \frac{G^{-1}}{2} + \frac{H}{4} (S+S^T)
    \Big)\lambda+\frac{3}{4} HG\:\tr (S+S^T)\bigg\}\, .
\end{equation}

\section{The effective $\sigma$-model}
\label{sec:sigma}

The normalization $Z$ can be interpreted as the generating functional for a theory defined by the action ${\cal S}(U)$
\begin{equation}
    Z=\TR\varrho =  \int {\cal D}U e^{- {\cal S}(U)}\, ,
\end{equation}
where
\begin{equation}
    {\cal S}(U) = \frac{M}{4} \lambda \lambda
    + \frac{1}{8} \lambda \:H (S+S^T)\:\lambda
    - \frac{1}{4\pi^2} HM^2 \tr S\, .
\end{equation}

We simplify this expression using
\begin{eqnarray}
    \label{eq:ll}
   \lambda \lambda  &=& \frac{2}{g^2}\tr( \partial U\partial U^\dagger)\, , \\
    \label{eq:lr}
  \lambda S^T \lambda &= &\lambda S \lambda
    =- \frac{1}{2 g^2}\tr\Bigl[(U^\dagger \partial U - \partial U^\dagger U)
    (\partial U U^\dagger - U\partial U^\dagger) \Bigr] \, ,\\
    \label{eq:trs}
   \tr S &=& \tr S^T = \tr U^\dagger \tr U -1\, .
\end{eqnarray}

Inserting these into the action we get
\begin{equation}
    \label{eq:action}
    {\cal S}(U) ={M\over 2 g^2} \tr( \partial U\partial U^\dagger)
    - \frac{H}{8 g^2}\tr\Big[(U^\dagger \partial U - \partial U^\dagger U)
    (\partial U U^\dagger - U\partial U^\dagger) \Big]
    - \frac{1}{4\pi^2} H M^2 \tr U^\dagger \tr U\, ,
\end{equation}
where $U$ independent pieces have been dropped.

As noted before, this effective $\sigma$-model is nothing but the
effective theory for the low momentum modes of the Polyakov loop
variable. Its status and applicability region are different from
the usual perturbative effective actions, see e.g. \cite{chris}.
The standard effective action is calculated in perturbation theory
and is valid at high temperature. Our effective action
eq.~(\ref{eq:action}) depends on the variational parameters $M$
and $H$, and in a sense is a variational effective action. Also
due to our restrictions to small values of $H$, a priori we do not
expect it to be valid at high temperatures but, it rather, should
represent correctly the physics in the phase transition region.

Another important difference is that our effective $\sigma$-model
does not have the local gauge invariance $U(x)\rightarrow
V^\dagger(x)U(x)V(x)$ which is usually associated with the
effective action for the Polyakov loop. The reason for this is
that our setup is different from that of the standard finite
temperature calculation. The way this gauge invariance usually
appears
 is the following. Consider the calculation of any gauge
invariant observable
\begin{equation}
\langle {\cal O}\rangle=\int {\cal D}U\TR[\exp\{-\beta {\mathsf H}\}{\cal O}g(U)]\, ,
\end{equation}
where $g(U)$ is the quantum mechanical operator of the gauge
transformation represented by the matrix $U$. This expression for
fixed $U$ can be compared to the same expression but with $U$
gauge transformed
\begin{equation}
{\rm Tr}[\exp\{-\beta {\mathsf H}\}{\cal O}g(V^\dagger UV)]={\rm
Tr}[\exp\{-\beta {\mathsf H}\}{\cal O}g(V^\dagger) g(U)g(V)]={\rm
Tr}[\exp\{-\beta {\mathsf H}\}{\cal O}g(U)]\, .
\end{equation}
The last equality here follows from the fact that both ${\cal O}$ and
$\exp\{\beta {\mathsf H}\}$ are gauge invariant, and thus the
operator $g(V^\dagger)$ can be commuted all the way to the left.
The only effect of the transformation is then to change the basis
over which the trace is being taken, which obviously leaves the
trace invariant.

Our setup is somewhat different. Expectation values are calculated
as
\begin{equation}
\int {\cal D}U {\rm Tr}[\tilde\varrho g(U) {\cal O}]
\end{equation}
with $\tilde\varrho$ defined in eq.~(\ref{tilderho}). This
expression is altogether gauge invariant, since the integral over
$U$ correctly projects only the contribution of gauge singlet
states. However the operator $\tilde\varrho$ is not itself
explicitly gauge invariant. For that reason the gauge
transformation operator $g(V^\dagger)$ cannot be commuted through
it, and thus
\begin{equation}
{\rm Tr}[\tilde\varrho {\cal O}g(V^\dagger UV)]\ne{\rm Tr}[\tilde\varrho
{\cal O}g(U)]
\end{equation}
even for gauge invariant operators ${\cal O}$. This manifests itself as
absence of local gauge invariance in the action of the effective
$\sigma$-model, eq.~(\ref{eq:action}).

Nevertheless, we stress again that the meaning of the $SU(N)$
valued field $U$ is precisely the same as that of the Polyakov
loop.

\section{The Calculation of the Free Energy}
\label{sec:free}

To find the best variational density matrix we have to minimize
the free energy with respect to the variational parameters $M$ and
$H$. The Helmholtz free energy ${\sf F}$ of the density matrix
$\varrho$ is given by
\begin{equation}
    {\mathsf F} = \langle {\mathsf H} \rangle
     - {\mathsf T}  {\mathsf S}\, ,
    \end{equation}
where ${\sf H}$ is the standard Yang-Mills Hamiltonian
eq.~(\ref{ham}), ${\sf S}$ is the entropy, and ${\sf T}$ is the
temperature.

Thus
\begin{equation}
\label{eq:freeenergy}
     {\mathsf F}
    = \frac{1}{2}\Big(
    \TR ({E}^2\varrho)
    +\TR ({B}^2\varrho)
    \Big)
    + {\mathsf T}\cdot\TR( \varrho \ln\varrho) \, .
\end{equation}
First of all we need to perform the integration over the gauge
fields, and reduce this expression to the average of a
$U$-dependent operator in the effective $\sigma$-model. In fact,
as we shall see soon, to leading order in $H$ the only nontrivial
calculation we need to perform is that of the entropy.

We will calculate the entropy up to the first nontrivial order in
$H$. As we now show, the leading term at small $H$ is $O(H\ln H)$.

Let us denote by $\varrho_{0}$ the density matrix of the pure
state with $H=0$:
\begin{equation}
    \varrho_{0} = |0\rangle\langle 0|\, .
\end{equation}
Here $|0\rangle$ does not denote necessarily the actual ground state,
but rather a projected Gaussian state with arbitrary $M$.  Now,
since the matrix elements of the density matrix can be expanded in
powers of $H$, to leading order we can write
\begin{equation}
    \varrho = \varrho_{0} + \delta\varrho\, ,
\end{equation}
where $\delta\varrho$ is $O(H)$.

Imagine that we have diagonalized $\varrho$. It will have one
large eigenvalue $\alpha_{0} = 1 -O(H)$, which corresponds to the
eigenstate
\begin{equation}
    |0'\rangle = |0\rangle +O(H)\, .
\end{equation}
All the rest of the eigenvalues $\alpha_{i}$ are at most  $O(H)$.
Then the entropy can be written as
\begin{equation}
    \label{eq:entropy1}
    {\mathsf S}= - \TR(\varrho\ln\varrho)
    = -\alpha_{0}\ln\alpha_{0} - \sum_{i=1}^{\infty}\alpha_{i}\ln\alpha_{i}\, .
\end{equation}

The second term is $O(H\ln H)$, and it is the coefficient of this
term that we will now calculate. Neglecting $O(H)$ corrections,
we can substitute $\alpha_{i}=H/M$ under the logarithm. Thus to
leading logarithmic order
\begin{equation}
    {\mathsf S} = -  \sum_{i}\alpha_{i}\ln H/M\, .
\end{equation}
So that all we have to calculate is $\sum_{i}\alpha_{i}$. Let
\begin{equation}
    |0'\rangle = |0\rangle + H |x\rangle\, .
\end{equation}
Then
\begin{equation}
    \varrho = \alpha_{0} |0'\rangle\langle 0'|
    + \sum_{i=1}^{\infty}\alpha_{i} |x_i\rangle\langle x_i|
\end{equation}
with
\begin{equation}
    \langle x_{i}|0'\rangle = 0\, .
\end{equation}

Note that $\langle 0|x\rangle \neq 0$, but
\begin{equation}
    \langle 0|x\rangle + \langle x|0\rangle = 0\, ,
\end{equation}
since $|0'\rangle$ has to be normalized at $O(H)$. Also
\begin{equation}
    \langle x_{i}|0\rangle + H\langle x_{i}|x\rangle = 0\, .
\end{equation}
Thus the overlap $\langle x_{i}|0\rangle$ is $O(H)$, and we have
\begin{equation}
    \varrho = \alpha_{0} |0\rangle\langle 0|
    + H \Big( |0\rangle\langle x| + |x\rangle\langle 0|\Big)
    + \alpha_{i} |x_{i}\rangle\langle x_{i}|\, .
\end{equation}
Multiplying this by $\varrho_{0}$ we get
\begin{eqnarray}
    \varrho_{0}\cdot\varrho
    &= &\alpha_{0}\cdot\varrho_{0} +H |0\rangle\langle x|
    + H \langle 0|x \rangle |0\rangle\langle 0| \, ,\\
    \varrho\cdot\varrho_{0}
    &=& \alpha_{0}\cdot\varrho_{0} +H |x\rangle\langle 0|
    + H \langle x|0 \rangle |0\rangle\langle 0|\, .\nonumber
\end{eqnarray}
Thus,
\begin{equation}
    \varrho_{0}\cdot\varrho + \varrho\cdot\varrho_{0} - \varrho =
    \alpha_{0}\cdot\varrho_{0} - \alpha_{i} |x_{i}\rangle\langle x_{i}|\, .
\end{equation}
Multiplying by $\varrho_{0}$ again, we get rid of
$|x_{i}\rangle\langle x_{i}|$ to $O(H)$
\begin{equation}
    \alpha_{0}\cdot\varrho_{0}=
    \varrho_{0}\cdot\varrho
    + \varrho_{0}\cdot\varrho\cdot\varrho_{0}
    - \varrho_{0}\cdot\varrho =
    \varrho_{0}\cdot\varrho\cdot\varrho_{0}\, .
\end{equation}
Then,
\begin{equation}
    \alpha_{0}=
    \TR(\varrho_{0}\cdot\varrho)\, .
\end{equation}
Since $\TR\varrho =1$ we have
\begin{equation}
    \sum_{i}\alpha_{i} = 1 -\alpha_{0} = \TR(\varrho_{0}(1-\varrho))
\end{equation}
which, inserted into eq.~(\ref{eq:entropy1}), gives
\begin{equation}
    {\mathsf S}= - (1-\TR(\varrho_{0}\cdot\varrho))\ln H/M \, .
\end{equation}

The derivation has been given for the normalized density matrices
$\varrho_{0}$ and $\varrho$. In terms of our Gaussian matrices we
should restore the normalization factors $Z$ and $Z_0$, so that
finally we have
\begin{equation}
    {\mathsf S}
    =\Big(
    \frac{\TR(\varrho_{0}\cdot\varrho)}{\TR\varrho_{0}\cdot\TR\varrho}-1
    \Big) \ln H/M \, .
\end{equation}
It is easy to check that to $O(H)$
\begin{equation}
    \TR(\varrho_{0}\cdot\varrho) = (\TR\varrho_{0})^{2}\, .
\end{equation}
Then
\begin{equation}
    {\mathsf S}
    =\Big(
    \frac{\TR\varrho_{0}}{\TR\varrho}-1
    \Big) \ln H/M \, .
\end{equation}

From eq.~(\ref{eq:trvarrho}), it is clear that
\begin{equation}
    \TR\varrho =
    \bigg[
    1+ H\Big(
    \frac{1}{4\pi^2}M^2\tr S
    -\frac{1}{4}\lambda S \lambda
    \Big)
    \bigg] \cdot \TR\varrho_{0}
 \end{equation}
Using eqs.~(\ref{eq:lr},\ref{eq:trs}) we finally get
\begin{equation}
    {\mathsf S}
    = -\biggl[<
    \frac{1}{8 g^{2}}
    \tr (U^\dagger \partial U - \partial U^\dagger U)
    (\partial U U^\dagger - U\partial U^\dagger)
    + \frac{1}{4\pi^2}M^2 (\tr U^{\dagger}\tr U -1)>_U
    \biggr]
    H\ln H/M\, .
    \label{entr}
\end{equation}

To leading order in $H$, the averaging over $U$ in this expression
has to be performed with the $\sigma$-model action at vanishing
$H$.

Expression eq.~(\ref{entr}) has the following striking property.
For $M<M_c$ it vanishes identically. The reason is very simple.
The first term in eq.~(\ref{entr}) is the product of the left
handed $SU(N)$ current and the right handed $SU(N)$ current in the
$\sigma$-model. Thus it transforms as an adjoint representation
under each one of the $SU(N)$ factors of the $SU_L(N)\otimes
SU_R(N)$ transformation. The same is also true for the second term
in eq.~(\ref{entr}). The $\sigma$-model action itself at $H=0$
obviously is invariant under the whole $SU_L(N)\otimes SU_R(N)$
group. Now, at $M<M_c$, the symmetry group is not spontaneously
broken, and thus any operator which is not a scalar has a
vanishing expectation value. It follows immediately that the
entropy has an $O(H\ln H/M)$ contribution only for $M>M_c$, when
the $SU_L(N)\otimes SU_R(N)$ group is spontaneously broken down to
$SU_V(N)$.

This observation makes our task considerably simpler. Since for
$M<M_c$ the entropy is zero, we do not have to consider at all the
disordered phase of the effective $\sigma$-model. In this
disordered phase the free energy coincides with energy, and thus
the calculation is identical to the calculation at zero
temperature. There is one slight subtlety here.  We expect of
course that since the energy alone  must have a minimum on a pure
state, the nonvanishing $H$ should give always a positive
contribution to the energy. Thus in the absence of the entropy
contribution, $H$ should vanish in the minimal energy state.
Nevertheless since in \cite{kk} the energy at $\mathsf{T}=0$ was
only minimized with respect to $M$, we have to check this property
explicitly. We have indeed performed this check and found that for
small $H$ the derivative of the expectation value of the energy
with respect to $H$ is strictly positive.

 Thus we only need to consider the effective
$\sigma$-model in the ordered phase. As at $\mathsf{T}=0$ we
perform the calculations in the ordered phase to leading order in
$\alpha_s$. Since there are no $O(H\ln H/M)$ corrections to energy
at this order, the result for the energy in the disordered phase
is identical to the result at zero temperature,
eq.~(\ref{largeM}). Thus our expression for the free energy is
\begin{equation}
{\mathsf F}={N^2-1\over 120\pi^2}M^4+{\mathsf T}\biggl(\langle
    \frac{1}{8 g^{2}}
    \tr (U^\dagger \partial U - \partial U^\dagger U)
    (\partial U U^\dagger - U\partial U^\dagger)
    + \frac{1}{4\pi^2}M^2 (\tr U^{\dagger}\tr U -1)\rangle_U
    \biggr)
    H\ln H/M\, .
\end{equation}

We now average over $U$ in the leading order perturbation theory.

\section{The $\sigma$-model perturbation theory --- minimization of the free energy
and the Debye mass}

We parameterize  the $U$ matrices as

\begin{equation}
    U =    \exp\Big\{-\frac{i}{2} \phi^{a}\tau^{a}\Big\}\, .
\end{equation}

Although we only need the leading order, it is instructive to
check that the order $g^{2}$ term in the expansion is indeed
small. To this order we have
\begin{equation}
    U \simeq
    \Big(1 - \frac{i}{2}\phi^{a}\tau^{a}
    - \frac{1}{8} \phi^{a}\phi^{b}\tau^{a}\tau^{b}
    + \frac{i}{48} \phi^{a}\phi^{b}\phi^{c}\tau^{a}\tau^{b}\tau^{c}
    \Big)\, .
\end{equation}
So that the $\sigma$-model action becomes
\begin{eqnarray}
\label{pertact}
    {\cal S} &=& {M\over 2g^2}\tr(\partial U \partial U^{\dagger})\\
    &=& \frac{M}{4g^2} \partial\phi\partial\phi
    + \frac{M}{192g^{2}} (\partial\phi^{a})(\partial\phi^{c})
    \phi^{b}\phi^{d}
   \tr \Big[
   \tau^{a}\tau^{b}\tau^{c}\tau^{d}
    -\tau^{a}\tau^{c}\tau^{b}\tau^{d}
     \Big]\, .\nonumber
\end{eqnarray}
The propagator of the phase field $\phi$ is thus
\begin{equation}
    \langle \phi^{a}\phi^{b} \rangle =
    \frac{2g^2}{M k^{2}} \delta^{ab}\, .
\end{equation}
To get the idea of the quality of this perturbative expansion we
can calculate for example $\langle {\cal S}\rangle$. In this calculation
one has to take into account the fact that the measure in the path
integral over the phase $\phi^a$ is not the simple ${\cal D}\phi$,
but rather the group invariant $U(N)$ measure $\mu$ . To first
order in $g^2$ it is
\begin{equation}
\mu={\cal D}\phi^a \exp\Big\{{M^3N\over 144\pi^2}\int d^3x \phi^2(x)\Big\}\, .
\label{measure}
\end{equation}
Taking this into account we find that $\langle {\cal S}\rangle$ gets no
correction of order $g^2$. We thus feel confident that the use of
the perturbation theory in the ordered phase of the $\sigma$-model
is an admissible approximation.
 In the following we will only keep leading order expressions.

Calculating to leading order the entropy eq.~(\ref{entr}) and
keeping only the $O(N^2)$ terms we find
\begin{equation}
    \langle {\mathsf S} \rangle
    =-\frac{N^{2}}{6\pi^{2}}M^{2} H\ln {H\over M}\, .
\end{equation}

Introducing the dimensionless quantity
\begin{equation}
    h=\frac{H}{M}
\end{equation}
we can write the expression for the free energy as
\begin{eqnarray}
     {\mathsf F}  &=&
    \langle {\mathsf H} \rangle
    -{\mathsf T}\langle {\mathsf S} \rangle \nonumber \\
    &=& \frac{N^{2}}{120\pi^{2}}
    M^{4}
    + {\mathsf T}\frac{N^{2}}{6\pi^{2}}M^{3}  h\ln  h\, .
\end{eqnarray}
We now have to minimize this expression with respect to $h$ and
$M$. It is convenient to first perform the minimization with
respect to $h$ at fixed $M$. This obviously gives
\begin{equation}
    \frac{\partial {\mathsf F}} {\partial h}=0
    \rightarrow {h}=\frac{1}{e}\, .
\end{equation}
Thus as a function of $M$ only, the free energy becomes
\begin{equation}
     {\mathsf F}
    = \frac{N^{2}}{120\pi^{2}}
    M^{4}
    - \frac{\mathsf T}{e}\frac{N^{2}}{6\pi^{2}}M^{3}\, .
\end{equation}
Now minimizing with respect to $M$ we find
\begin{equation}
    \frac{\partial {\mathsf F}} {\partial M}=0
    \rightarrow M = \frac{15 {\mathsf T}}{e}\, .
\end{equation}

Thus for $M\ge M_c$ the free energy of the best variational
density matrix as a function of temperature is
\begin{equation}
    {\mathsf F}_{M\ge M_c}
    = -\frac{N^{2}}{360\pi^{2}}
    \Big(\frac{15 {\mathsf T}}{e}\Big)^{4}\, .
    \label{freeen}
\end{equation}

We now have to compare this value with the free energy for $M\le
M_c$. As we have discussed above, this is given by the expectation
value of the Hamiltonian alone, and is minimized at $M=M_c$. Its
value is
\begin{equation}
{\mathsf F}_{M\le M_c}=-{N^2\over 30\pi^2}M_c^4\, .
\end{equation}
Comparing the two expressions we find
\begin{equation}
\mathsf{T_{c}}= \frac{12^{\frac{1}{4}} e}{15} M_c\, .
\end{equation}
Using the value of $M_c$ from eq.~(\ref{mc}) we have
\begin{equation}
\mathsf{T_{c}}= 450 \mathrm{Mev}\, .
\end{equation}

For $\mathsf{T}\le \mathsf{T_{c}}$ the free energy is minimized in
the variational state with $M=M_c$. In our approximation this
state is the same as at zero temperature. Its entropy vanishes,
and the effective $\sigma$-model is in the disordered phase. The
Polyakov loop vanishes, $\langle U\rangle=0$ and according to the
standard wisdom this is a confining state.

For $\mathsf{T}\ge \mathsf{T_{c}}$ the best variational state is very different. The
entropy of this state is nonzero,
\begin{equation}
 {\mathsf S}={N^2\over 6\pi^2 e}\Big({15 {\mathsf T}\over e}\Big)^3\, .
\end{equation}
The Polyakov loop is nonzero $\langle U\rangle\ne 0$ and thus the high
temperature density matrix describes a deconfined phase.

Finally, we note that in the deconfined phase our best variational
density matrix has a nonvanishing ``electric screening" or ``Debye"
mass. The Debye mass is conveniently defined as the ``mass" of the
phase of the Polyakov loop. This mass is nonvanishing in our
calculation for the following reason. As long as $H=0$, the
effective $\sigma$-model action has a global $SU_L(N)\otimes
SU_R(N)$ symmetry. Thus in the ordered phase of the $\sigma$ model
the phases $\phi^a$ are massless. However, as discusses above, the
terms of order $H$ in eq.(\ref{eq:action}) break this symmetry
explicitly down to the diagonal $SU_V(N)$. As a result the would
be ``Goldstone" phases $\phi^a$ acquire mass. To calculate this
mass it is convenient first to note that to $O(g^2)$
\begin{equation}
\tr(U^\dagger \partial U - \partial U^\dagger U)
    (\partial U U^\dagger - U\partial U^\dagger)=-4\tr\partial U^\dagger
    \partial U
    -{1\over
    4}\phi^a\phi^c\partial\phi^b\partial\phi^d
    \tr\big(\tau^a\tau^b\tau^c\tau^d-\tau^a\tau^c\tau^b\tau^d\big)\, .
    \label{interact}
\end{equation}
The contribution of the $SU_L(N)\otimes SU_R(N)$ term to the mass
cancels against the contribution of the measure
eq.(\ref{measure}).
 Using
eqs.(\ref{eq:action}, \ref{interact}) we then find to $O(g^2)$ and
to leading order in $H$
\begin{equation}
M_D^2={4\over 3\pi}\alpha_s(M)NMH \, .
\end{equation}

As a function of temperature we have
\begin{equation}
M_D^2=\alpha_s\Big({15\over e}T\Big)N{300\over \pi e^3}{\mathsf T}^2\, .
\label{md}
\end{equation}

\section{Discussion}
Let us summarize the results of our variational calculation. We
find the phase transition at a temperature of about
$\mathsf{T_{c}}\simeq 450 {\rm Mev}$. The transition is strongly
first order at large $N$. The latent heat is $\Delta E={N^2\over
90\pi^2 }\big ({15 {\mathsf T}\over e}\big)^4$. Below the
transition the entropy is zero, the best variational state is the
same as at zero temperature, and the average value of the Polyakov
loop is zero. Above the transition, the entropy is nonzero and
proportional to the number of ``coloured" degrees of freedom,
${\mathsf S}\propto N^2$. The average value of the Polyakov loop
is nonzero and the phase is deconfined.

It is quite interesting that  at high temperature our formulae
numerically are quite close to the predictions of free gluon
plasma. In particular, our value for the free energy,
eq.~(\ref{freeen}), should be compared to the free gluon plasma
expression
\begin{equation}
{\mathsf F}_{free}= -\frac{(N^2-1)\pi^2}{45} {\mathsf T}^4\, .
\end{equation}
The ratio between the two is
\begin{equation}
\frac{{\mathsf F}_{free}}{{\mathsf F}_{var}}\simeq 0.85 \,
.\label{ratio}
\end{equation}
The ratio of the entropies is the same.

Interestingly we get the same ratio  comparing our value for the
Debye mass eq.(\ref{md}) with the leading order perturbative one,
$M_{pert}^2={4\pi\over 3}\alpha_sNT^2$,
\begin{equation}
{M_{pert}^2\over M_D^2}\simeq 0.85 \, .\label{ratio1}
\end{equation}

The pressure approaches its asymptotic value according to the
simple formula
\begin{equation}
\frac{{\mathsf P}({\mathsf T})}{{\mathsf P}_{\rm asympt}}=1-\frac
{\mathsf{T_{c}}^4}{{\mathsf T}^4}\, .
\end{equation}

One has to take the comparison eqs.(\ref{ratio},\ref{ratio1}) with
a grain of salt. As explained above, our calculations were
performed assuming small $H$. A priori we expect that this
restriction should confine us to not too large temperatures.
Interestingly, however, the minimization of the free energy
resulted in the value $H/M=1/e$ independently of temperature.
Thus, we feel ourselves justified to consider the comparison
eq.(\ref{ratio}) as meaningful.

The main features of these results are indeed what we expect from
the deconfinement phase transition on general grounds. It is
extremely gratifying that a simple minded calculation such as ours
 does qualitatively so well in such a complicated problem.
It therefore appears that the projection of the trial density
matrix on the gauge invariant Hilbert space is, just like at zero
temperature, the crucial feature that dictates most if not all the
important aspects of the low energy and low temperature physics.
In the context of the present calculation the most important
effect of the gauge projection is obviously vanishing of the
entropy in the low temperature phase. We stress that this feature
was not at all built into our initial ansatz, but followed
naturally and unavoidably in the disordered phase of the effective
$\sigma$-model.

 Quantitatively, our calculation of course should be taken for what
 it is --- an approximate implementation of the variational
 principle. The projection over the gauge group, which as we saw
 is so physically important, is what makes the calculational task
 difficult. The most severe simplifications that we had to impose
 are the perturbation theory in the ordered phase of the $\sigma$-model
 and the assumption of smallness of $H$.

We believe that both these approximations  affected the quality of
our results. In particular, in the leading order of the
perturbation theory the expectation value of the Polyakov loop $U$
is equal to unity. The actual value of $U$ on the ordered side of
the transition according to \cite{kogut} is close to one half.
Thus our perturbative calculation is rather more reliable somewhat
further away from the transition. In line with this we expect that
the estimate for the critical temperature we obtained here is
somewhat higher than we would get, had we treated the
$\sigma$-model more accurately in the transition region. This is
consistent with the fact that our result for $\mathsf{T_{c}}$ is
by about $50\%$ higher than the lattice value of $270 {\rm Mev}$.

The smallness of $h$ is also quite important. The value of $h=1/e$
that we obtain is in fact a reasonably small number, so omitting
the corrections in powers of $h$ is fairly safe. On the other
hand, the terms linear in $h$ but not enhanced by $\ln h$, which
we have ignored in the present calculation, have to be accounted
for in a more careful way. Once these terms are taken care of our
calculation can be extended to high temperatures.

 We believe that
 with some effort both these limitations can be overcome, at least
 to some extent. It should be possible to treat the $\sigma$-model
 action in a better way, perhaps along the lines of a
 continuum version of \cite{kogut}.
As for terms linear in $h$, those should be accessible in
perturbative expansion in $h$, once the nonanalytic $O(h \ln h)$
terms have been understood. We plan to continue investigations
along these lines and hope to improve on the quality of our
results.

Notwithstanding this critique, we are very much encouraged by the
results of this exploratory investigation. We believe that a
reasonably simple improvement on the calculational methods has a
good chance to significantly improve the results and bring them
into a better quantitative agreement with the lattice data.

\begin{acknowledgments}
We thank R. Buffa and J.P. Garrahan for their contribution during
the early stages of this work. I.K. and A.K. thank the Royal
Society for the joint project RS-CNRS grant. J.G.M. thanks the
Department of Mathematics and Statistics, University of Plymouth,
for hospitality while part of this work was done.
\end{acknowledgments}


\begin{thebibliography}{99}
\bibitem{note1} At first sight it may seem
surprising that the standard action for Polyakov loop is invariant
under the local gauge transformation $P\rightarrow V^\dagger P V$,
whereas the action of our effective $\sigma$ model does not
possess this gauge invariance. We will explain this point later
when discussing our setup at finite temperature.
\end{thebibliography}
\end{document}